\documentclass[sigconf,natbib=false]{acmart}
\AtBeginDocument{%
  }

\setcopyright{acmlicensed}
\copyrightyear{2018}
\acmYear{2018}
\acmDOI{XXXXXXX.XXXXXXX}
\acmConference[Conference acronym 'XX]{Make sure to enter the correct
  conference title from your rights confirmation email}{June 03--05,
  2018}{Woodstock, NY}
\acmISBN{978-1-4503-XXXX-X/2018/06}

\RequirePackage[
  datamodel=acmdatamodel,
  style=acmnumeric,
  ]{biblatex}

\usepackage{array}
\usepackage{tabularx}
\usepackage{placeins}
\usepackage{float}
\newcolumntype{L}{>{\raggedright\arraybackslash}X}

\begin{document}
\def\sysname{{UniSGR }}
\title{UniSGR: Unified Framework for Semantic ID Generation and Ranking}

\author{Jiawei Sun}
\email{xiangxing.sjw@alibaba-inc.com}
\authornote{These authors contributed equally to this research.}
\orcid{0009-0001-6018-234X}
\affiliation{%
  \institution{Alibaba International Digital Commerce Group}
  \city{Beijing}
  \country{China}
}

\author{Jun Yang}
\authornotemark[1]
\email{yj494343@alibaba-inc.com}
\affiliation{%
  \institution{Alibaba International Digital Commerce Group}
  \city{Beijing}
  \country{China}
}

\author{Ziyue Guo}
\email{guoziyue.gzy@alibaba-inc.com}
\affiliation{%
  \institution{Alibaba International Digital Commerce Group}
  \city{Beijing}
  \country{China}
}

\author{Dongyue Xu}
\email{xudongyue.xdy@alibaba-inc.com}
\affiliation{%
  \institution{Alibaba International Digital Commerce Group}
  \city{Beijing}
  \country{China}
}

\author{Jianan Yan}
\authornote{Corresponding author.}
\email{jianan.yjn@alibaba-inc.com}
\affiliation{%
  \institution{Alibaba International Digital Commerce Group}
  \city{Beijing}
  \country{China}
}

\author{Lifang Deng}
\authornotemark[2]
\email{wanmei.dlf@alibaba-inc.com}
\affiliation{%
  \institution{Alibaba International Digital Commerce Group}
  \city{Beijing}
  \country{China}
}

\author{Xiaoyi Zeng}
\email{yuanhan@taobao.com}
\affiliation{%
  \institution{Alibaba International Digital Commerce Group}
  \city{Hangzhou}
  \country{China}
}

\renewcommand{\shortauthors}{Sun et al.}

\begin{abstract}
Recommendation systems play a pivotal role in modern e-commerce platforms. While generative retrieval has emerged as a promising paradigm for alleviating the limitations of multi-stage cascade architectures, existing methods still struggle with fine-grained multi-objective ranking. To bridge this gap, we propose UniSGR, a \textbf{Uni}fied framework for \textbf{S}emantic ID \textbf{G}eneration and \textbf{R}anking. UniSGR adopts a two-stage training paradigm: a multi-scenario pre-training stage that learns from mixed business-scenario data, followed by a scenario-specific alignment stage that jointly optimizes Value-Aware Parallel Multi-Token Prediction (VA-PMTP) and a unified multi-objective ranking module. To better align generation with downstream ranking, we introduce Task-Aware Tokens (TAT) guided by Funnel-Aware Contrastive Learning. Furthermore, we propose Semantic Tree Attention with Reorganized KV cache (STARK), an inference strategy that removes key efficiency bottlenecks in conventional beam search. Extensive offline experiments on a large-scale e-commerce platform demonstrate the effectiveness and scalability of UniSGR.
\end{abstract}



\maketitle

\section{Introduction}

Recommendation systems (RSs) are indispensable to modern e-commerce platforms~\cite{alamdari2020systematic}. Industrial scale systems are typically organized as cascaded pipelines, where candidate retrieval is followed by pre-ranking, ranking, and re-ranking stages~\cite{isinkaye2015recommendation,ko2022survey}. In the DLRM era~\cite{mudigere2022software,covington2016deep,zhou2018deep}, retrieval is commonly implemented with efficient dual-tower matching models: user and item features are encoded independently, item embeddings are indexed offline, and large-scale maximum inner-product search retrieves candidates under strict latency constraints.

Despite its effectiveness, this cascade architecture optimizes each module for a local subtask rather than the end-to-end recommendation objective. Downstream rankers can only reorder candidates retained by upstream retrieval, so relevant items filtered out early cannot be recovered later. This stage-wise decomposition limits the possibility of truly end-to-end recommendation.

Generative retrieval has recently emerged as a promising alternative~\cite{deng2025onerec,zhai2024actions}. By formulating recommendation as autoregressive generation, these methods directly generate item identifiers, often represented as discrete semantic IDs, and thus provide a more unified view of candidate generation.

However, existing generative retrieval methods still have limited ranking capability. They can select semantically relevant candidates, but often lack the fine-grained discrimination required by industrial ranking. A straightforward solution is to add a discriminative ranker after generative retrieval, but this reintroduces objective mismatch: the generator is optimized to produce plausible candidates, whereas the ranker optimizes final utility over a truncated candidate set.

The training paradigm is another key factor. Large-scale platforms contain multiple business scenarios with heterogeneous intents, behaviors, and objectives. Training exclusively on a target scenario limits the transferability of learned knowledge, while naively pooling all data may weaken scenario-specific adaptation. We therefore adopt multi-scenario pre-training followed by scenario-specific alignment, a paradigm analogous to broad pre-training and downstream adaptation in foundation models~\cite{brown2020language,devlin2019bert}.

Motivated by this perspective, we propose \sysname, a \textbf{Uni}fied framework for \textbf{S}emantic ID \textbf{G}eneration and \textbf{R}anking. \sysname couples generative retrieval and discriminative ranking in a single learning framework, incorporating ranking-oriented supervision into semantic ID generation so that generated candidates are better aligned with downstream business objectives.

The framework further supports multi-objective optimization, enabling click, add-to-cart, and purchase signals to be incorporated into one training process. As a result, semantic ID generation and ranking are optimized toward a shared recommendation goal.

Our major contributions can be summarized as follows:
\begin{enumerate}
    \item We propose UniSGR, a unified framework that seamlessly integrates semantic ID generation and multi-objective ranking, mitigating the gap inherent in traditional cascade architectures.
\item We design a two-stage training paradigm that combines multi-scenario pre-training with scenario-specific alignment, where Value-Aware Parallel Multi-Token Prediction (VA-PMTP) and Task-Aware Tokens (TAT) help align generation with business objectives.
\item We introduce STARK, an efficient tree-attention inference strategy tailored for semantic IDs, achieving a $200\%$ throughput improvement in industrial-scale scenarios.
\item Extensive offline evaluations and online A/B testing on a real-world e-commerce platform validate the effectiveness and scalability of UniSGR.
\end{enumerate}

\section{Related Work}

\subsection{Deep Learning Recommendation Model}

Traditional DLRMs rely on ID-centric sparse features and continuous dense features, followed by explicit or learned feature interactions. Wide \& Deep~\cite{cheng2016wide} combines memorization and generalization, while DeepFM~\cite{guo2017deepfm} jointly models low- and high-order feature interactions with shared embeddings. Attention-based models such as AutoInt~\cite{song2019autoint} and DIN~\cite{zhou2018deep} further capture diverse user interests from historical behaviors. DLRM~\cite{naumov2019deep} addresses system-level constraints between dense computation and sparse embedding lookup, and LONGER~\cite{chai2025longer} improves long-sequence modeling by decoupling retrieval and content representation spaces.

\subsection{Generative Recommendation Model}
Recent works leverage semantic representations and generative models to address the limitations of standard ID-based recommendation. TIGER~\cite{rajput2023recommender} introduces semantic tuple IDs via RQ-VAE~\cite{lee2022autoregressive}. Meanwhile, HLLM~\cite{chen2024hllm} integrates item-level semantic modeling with user-sequence modeling. To handle high-cardinality settings, HSTU~\cite{zhai2024actions} improves efficiency using linear gated aggregation. Furthermore, MTGR~\cite{han2025mtgr} investigates scaling behaviors in industrial generative recommendation. OneRec~\cite{deng2025onerec} and OneRec-Think~\cite{liu2025onerec} also explore end-to-end generative paradigms, incorporating preference optimization alongside reasoning-enhanced generation.

Unifying retrieval and ranking has also attracted increasing attention. OneRanker~\cite{oneranker}, GRank~\cite{Sun2025GRankTT}, and GPR~\cite{zhang2026gpr} reduce objective mismatch by coupling generation, ranking, or pre-training in industrial recommendation settings. Building on this direction, \sysname focuses on semantic ID-based generative recommendation and introduces a multi-objective ranking module directly on top of the generative decoder to align candidate generation with final business values.

\section{Methodology} 

\subsection{Data Arrangement}
The training data are organized according to the two-stage learning paradigm. In multi-scenario pre-training, we mix user behavior logs from multiple business scenarios and construct next-item generation samples from chronological interaction sequences. Each sample consists of user profile features, scenario/context features, previously interacted items, and the semantic ID sequence of the next interacted item. This stage is designed to expose the model to diverse user interests and item semantics spanning multiple scenarios.

In scenario-specific alignment, samples are derived from the target scenario and incorporate richer behavior labels. For a user session, we collect candidate target items associated with different behavior types, such as click, add-to-cart, and purchase. These behavior-specific targets are used in two ways: VA-PMTP treats them as generation targets with different business-value weights, while the ranking module predicts their corresponding behavior labels using shared generation representations. This arrangement enables generation and ranking to be jointly optimized on the same semantic ID representation space.

For serving, the input follows the same feature organization as training. The encoder consumes user profile, behavior sequence, and context features, while the decoder generates semantic IDs for candidate items. The ranking module then scores the generated candidates without relying on a separate item-ID based ranking model.

\subsection{Problem Formulation}

We formulate sequential recommendation as unified generative retrieval and multi-objective ranking. Let $\mathcal{U}$ and $\mathcal{I}$ denote the user and item sets. For each user $u$, the system observes user profile features $\mathbf{x}_u$ and a chronological interaction sequence $\mathcal{S}_u = [v_1, v_2, \ldots, v_L]$, where $v_t \in \mathcal{I}$. Each item $v$ is mapped to a discrete semantic ID sequence $\mathbf{s}_v = (c_{v,1}, c_{v,2}, \dots, c_{v,d})$, where $d$ is the number of semantic layers.

Our framework tackles this problem by aligning with the two-stage learning paradigm:

\textbf{1. Multi-Scenario Pre-training:} The model learns generalizable user interests from mixed business-scenario data and acts as a pure generative retriever:
\begin{equation}
P_\theta(v_{L+1} \mid \mathbf{x}_u, \mathcal{S}_u) = \prod_{j=1}^{d} P_\theta(c_{v_{L+1}, j} \mid \mathbf{x}_u, \mathcal{S}_u, c_{v_{L+1}, <j}).
\label{eq:generative_retrieval}
\end{equation}

\textbf{2. Scenario-Specific Alignment:} The pre-trained model is adapted to a target scenario by jointly optimizing value-aware next-token prediction and multi-objective ranking. Given semantic IDs and generative hidden states, the model predicts click, add-to-cart, and purchase labels with BCE losses:
\begin{equation}
\hat{y}_{v}^{(task)} = P_\phi(y^{(task)}=1 \mid \mathbf{x}_u, \mathcal{S}_u, \mathbf{s}_v), \quad task \in \{\text{click}, \text{atc}, \text{pay}\}.
\label{eq:multi_obj_ranking}
\end{equation}

\textbf{Inference Serving:} The aligned model decodes candidates via beam search and reuses the generated semantic IDs and hidden states to compute multi-objective scores. The serving system can then return re-ranked Top-$K$ recommendations without invoking a separate ranking model.

\subsection{Tokenizer}

We first derive multimodal collaborative item embeddings by fine-tuning Qwen3-VL with behavioral interaction pairs, then discretize them with RQ-VAE and Sinkhorn-Knopp assignment into a balanced 3-layer semantic codebook. 

\textbf{Collaborative-Aware Representation.} For item $i$, we feed textual metadata $T_i$ and visual features $V_i$ into Qwen3-VL, denoted as $f_\theta$, to obtain a multimodal embedding:
\begin{equation}
    e_i = f_\theta(T_i, V_i)
\end{equation}
We use an interacted target item $i^+$ as the positive sample, exposed-but-unclicked items $\mathcal{H}_i$ as hard negatives, and in-batch items $\mathcal{B}_i$ as easy negatives. The network is optimized with the following InfoNCE loss:
\begin{equation}
\mathcal{L}_{CL}
= -\log
\frac{\exp(\operatorname{sim}(e_i, e_{i^+}) / \tau)}
{\sum_{j \in \mathcal{C}_i} \exp(\operatorname{sim}(e_i, e_j) / \tau)} ,
\end{equation}
where $\mathcal{C}_i=\{i^+\}\cup\mathcal{H}_i\cup\mathcal{B}_i$, $\operatorname{sim}(x, y)$ denotes cosine similarity, and $\tau$ is the temperature hyperparameter.

\textbf{Quantization}. We apply Residual-Quantized VAE (RQ-VAE)~\cite{zeghidour2021soundstream} with Sinkhorn-Knopp balanced assignment, generating a 3-layer semantic hierarchy with codebook size $K=8192$.

\subsection{\sysname Framework}

\begin{figure*}[!htbp]
    \centering
    \includegraphics[width=0.9\textwidth]{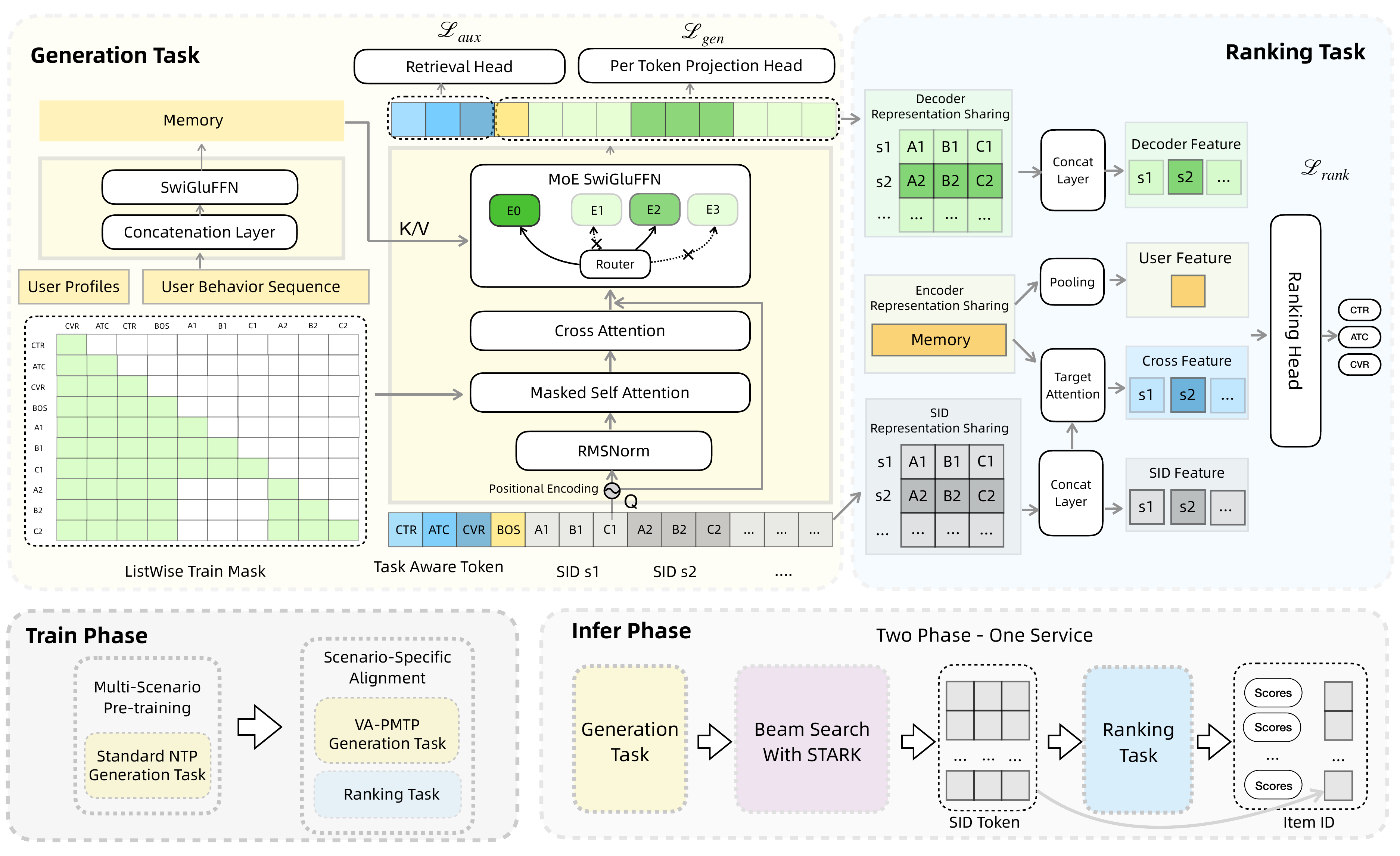}
    \caption{The overall architecture of UniSGR. It features a lightweight feature encoder for multimodal user behavior sequences and a sparse MoE decoder that unifies generative retrieval and multi-objective ranking using Task-Aware Tokens.}
    \label{fig:main}
\end{figure*}

Figure~\ref{fig:main} illustrates \sysname, an asymmetric Transformer-based encoder-decoder framework with a lightweight feature encoder and a deep sparse Mixture-of-Experts (MoE) decoder.

\textbf{Feature Encoder (MemoryNet)}. The encoder processes user profiles, previously interacted items, and contextual tags. Instead of self-attention, MemoryNet maps user profiles through an MLP and prepends them as prefix tokens to the behavior sequence. Each historical item is represented by its semantic ID embedding together with behavior and position information, so the encoder can distinguish both item identity and interaction type. Positional encoding, dropout, and a linear projection then produce a static memory representation $\mathbf{M}$ with linear complexity $\mathcal{O}(L)$. This design is deliberately lightweight: it retains the information required by the decoder while circumventing quadratic self-attention over long user histories.

\textbf{Sparse MoE Decoder}. The decoder autoregressively generates semantic ID tokens, initialized with a learnable BOS token and conditioned on $\mathbf{M}$ through cross-attention. A learned Token Type Embedding (TTE) is added to semantic ID embeddings to encode their hierarchical structure. This is critical because different semantic ID levels decodes distinct semantics: upper levels capture coarse semantic clusters, while lower levels provide finer-grained discrimination and item uniqueness. The decoder thus jointly learns sequence-level generation and level-aware semantic refinement.

For efficient and scalable decoding, we adopt Grouped Query Attention (GQA)~\cite{ainslie2023gqa} to reduce KV cache cost and replace standard FFNs with shared SwiGLU-MoE layers~\cite{dai2024deepseekmoe}. The shared expert captures common recommendation patterns, whereas routed experts specialize in distinct semantic subspaces and user-interest modes:
\[
y = \mathrm{Expert}_{\mathrm{shared}}(x) + \sum_{i \in \mathrm{TopK}} g_i(x)\,\mathrm{Expert}_i(x).
\]
We further employ RMSNorm~\cite{zhang2019root} and QK-Norm ~\cite{henry2020query} for stable training.

\subsection{Two-Stage Training Paradigm}
UniSGR adopts a two-stage training paradigm:
\begin{enumerate}
\item \textbf{Multi-scenario pre-training stage:} Captures general user interests and item semantics via Next Token Prediction on mixed data from multiple business scenarios.
\item \textbf{Scenario-specific alignment stage:} Introduces VA-PMTP and a unified multi-objective ranking module to align generation with target-scenario business goals.
\end{enumerate}

\subsubsection{Multi-Scenario Pre-Training Stage}
In contrast to training on sparse interactions from a single scenario, multi-scenario pre-training exposes UniSGR to diverse behavior signals aggregated across business scenarios. The pre-training employs the NTP loss:
\begin{equation}
\mathcal{L}_{\text{NTP}} = -\sum_{j=1}^{d} \log P_\theta(c_{v_{L+1}, j} \mid \mathbf{x}_u, \mathcal{S}_u, c_{v_{L+1}, <j})
\label{eq:ntp_loss}
\end{equation}
The resulting backbone serves as a robust foundation for subsequent scenario-specific alignment.

This stage initializes the semantic ID generator with broad cross-scenario interests, which is particularly beneficial for sparse target scenarios that otherwise suffer from poor coverage and overfitting.

\subsubsection{Value-Aware Parallel Multi-Token Prediction}
To capture multiple concurrent interests in a target scenario, we introduce Parallel MTP (PMTP), which uses a customized parallel mask for multiple autoregressive semantic-ID targets in the same session:
\begin{itemize}
    \item Within the same item: Apply causal masking between semantic IDs at all levels.
    \item Across different items: Enforce mutual invisibility.
    \item Shared BOS token: All semantic IDs can access the BOS token, unifying the representation of multiple interests through a single BOS.
\end{itemize}

We further introduce business-value weights over behaviors (e.g., purchase > add-to-cart > click > impression), yielding the VA-PMTP Loss:
\begin{equation}
\mathcal{L}_{\text{VA-PMTP}} = -\sum_{\tau} w_{\tau}\sum_{i=1}^{\mathcal{S}_{\tau}} \sum_{j=1}^{d} \log P_\theta(c_{v_{L+i}, j} \mid \mathbf{x}_u, \mathcal{S}_u, c_{v_{L+i}, <j})
\label{eq:vapmtp_loss}
\end{equation}
where $w_{\tau}$ is the business value weight and $\mathcal{S}_{\tau}$ is the item set for objective $\tau$ within the session.

Compared with standard NTP, VA-PMTP both increases target diversity via parallel behavior-specific targets and biases generation toward higher-value behaviors, yielding a candidate set more useful for downstream ranking.

\subsubsection{Unified Multi-Objective Ranking Module}
Scenario-specific alignment attaches a semantic ID-based multi-objective ranking module to the generative decoder. Rather than introducing an independent item ID-based ranker, this module reuses the semantic ID representations and decoder states produced during generation.

\textbf{Input Sharing}:
\begin{itemize}
    \item \textbf{Semantic ID Representation Sharing}: Generation and ranking share semantic ID representations; concatenated multi-level semantic IDs serve as candidate inputs to the ranking module.
    \item \textbf{Encoder Representation Sharing}: Generation and ranking share the encoder representations of user behavior sequences.
    \item \textbf{Decoder Representation Sharing (DRS)}: As the decoder proceeds with autoregressive decoding, semantic ID representations at each level progressively capture higher-order interactions between target items and user representations. These decoder states are reused by the ranking module.
\end{itemize}

\textbf{Model Architecture}:
\begin{itemize}
    \item \textbf{Target Attention}: Applies cross-attention over the behavior sequence encoder representations using candidate semantic IDs, enabling fine-grained user interest modeling.
    \item \textbf{PLE-based Multi-Objective Ranking}: User information, semantic ID representations, Target Attention outputs, and decoder representations are used as bottom-layer inputs to the Progressive Layered Extraction(PLE) module for multi-objective ranking.
\end{itemize}

This shared design enables ranking losses to propagate value-oriented gradients to the representations employed for generation.

\subsubsection{Task-Aware Tokens for Generation-Ranking Alignment}
A pure generation objective does not explicitly encode which business objective the representation should serve, while click, add-to-cart, and purchase prediction require related but different preference signals.

We therefore propose Task-Aware Tokens (TAT), which prepends learnable task tokens to the decoder input. This allows subsequent semantic ID representations to condition on objective-specific signals.

We introduce three learnable embedding vectors $\mathbf{e}_{\text{click}}$, $\mathbf{e}_{\text{atc}}$, and $\mathbf{e}_{\text{pay}}$, corresponding to the click, add-to-cart, and purchase objectives, respectively. They are prepended before the beginning-of-sequence token $\langle \text{bos} \rangle$, forming an augmented decoder input:
\begin{equation}
\mathbf{X}_{\text{dec}} = [\underbrace{\mathbf{e}_{\text{click}},\ \mathbf{e}_{\text{atc}},\ \mathbf{e}_{\text{pay}}}_{\text{task tokens}},\ \underbrace{\langle \text{bos} \rangle,\ s_1,\ s_2, \ldots}_{\text{sequence tokens}}]
\end{equation}

The sequence is processed under a unified causal mask: task tokens follow the click $\rightarrow$ add-to-cart $\rightarrow$ purchase funnel, and all semantic ID tokens can attend to preceding task tokens. Thus, objective-specific information is injected from the first decoding step.

Since task tokens occupy fixed prefix positions, TAT incurs no additional autoregressive steps during inference.

\subsubsection{Joint Training of Generation and Ranking}
The main training objective consists of two components---a generation task and a ranking task---which jointly drive end-to-end optimization.

\textbf{Generation Task}. Let $\mathbf{e}_{\text{task}} = (\mathbf{e}_{\text{click}}, \mathbf{e}_{\text{atc}}, \mathbf{e}_{\text{pay}})$ denote the task-token representations. The decoder performs VA-PMTP at the sequence token positions, using a cross-entropy loss:
\begin{equation}
\mathcal{L}_{\text{gen}} =
-\sum_{\tau} w_{\tau}\sum_{i=1}^{\mathcal{S}_{\tau}} \sum_{j=1}^{d}
\log P_{\theta}\!\left(
c_{v_{L+i}, j} \mid \mathbf{x}_u, \mathcal{S}_u, c_{v_{L+i}, <j}, \mathbf{e}_{\text{task}}
\right)
\end{equation}
Compared with the baseline without task tokens, this conditional probability includes $\mathbf{e}_{\text{task}}$, so generation is modulated by objective-specific preferences.

\textbf{Ranking Task}. Decoder outputs at $\langle \text{bos} \rangle, s_1, s_2, \ldots$ serve as task-aware features for click, add-to-cart, and purchase prediction.
The ranking loss is a weighted sum of the per-objective losses:
\begin{equation}
\mathcal{L}_{\text{rank}} = \sum_{\tau} \lambda_{\tau} \cdot \mathcal{L}_{\tau}(\hat{y}_{\tau}, y_{\tau})
\end{equation}
where $\mathcal{L}_{\tau}$ denotes the binary cross-entropy (BCE) loss for the corresponding objective.

\textbf{Joint Optimization}. The final training objective is:
\begin{equation}
\mathcal{L} = \mathcal{L}_{\text{gen}} + \alpha \cdot \mathcal{L}_{\text{rank}}
\end{equation}
where $\alpha$ is a balancing hyperparameter. The gradients from the ranking loss propagate back through the shared decoder parameters, further reinforcing the multi-objective awareness of the output representations at the sequence token positions.

\subsubsection{Funnel-Aware Contrastive Learning for Task Token Guidance}
Task tokens influence downstream sequence representations primarily through attention. Consequently, they are susceptible to weak supervision and semantic drift.

We introduce Funnel-Aware Contrastive Learning (FACL) as an auxiliary signal. For task token output $\mathbf{h}_{\tau}$, $\tau \in \{\text{click}, \text{atc}, \text{pay}\}$, we perform contrastive learning against item ID embeddings $\mathbf{v}_i$:
\begin{equation}
\mathcal{L}_{\tau}^{\text{aux}} = -\log \frac{\exp(\text{sim}(\mathbf{h}_{\tau},\ \mathbf{v}^+) / t)}{\exp(\text{sim}(\mathbf{h}_{\tau},\ \mathbf{v}^+) / t) + \sum_{j=1}^{K} \exp(\text{sim}(\mathbf{h}_{\tau},\ \mathbf{v}_j^-) / t)}
\end{equation}
where $\text{sim}(\cdot, \cdot)$ denotes cosine similarity and $t$ is a temperature. Positives follow the behavior associated with each task token, while negatives become progressively more challenging along the conversion funnel: random items for click, clicked-but-not-added items for add-to-cart, and clicked-but-not-purchased items for purchase.
\begin{itemize}
    \item \textbf{Click token}: positives are clicked items, and negatives are sampled from the item pool. This token captures coarse-grained interest.
    \item \textbf{Add-to-cart token}: positives are add-to-cart items, while negatives are clicked but not added items. This encourages the token to distinguish stronger purchase intent from click-level interest.
    \item \textbf{Purchase token}: positives are purchased items, while negatives are clicked but not purchased items. This guides the token toward the final conversion objective.
\end{itemize}

The auxiliary contrastive loss is:
\begin{equation}
\mathcal{L}_{\text{aux}} = \beta \cdot (\mathcal{L}_{\text{click}}^{\text{aux}} + \mathcal{L}_{\text{atc}}^{\text{aux}} + \mathcal{L}_{\text{pay}}^{\text{aux}})
\end{equation}

The complete training objective is:
\begin{equation}
\mathcal{L} = \underbrace{\mathcal{L}_{\text{gen}} + \alpha \cdot \mathcal{L}_{\text{rank}}}_{\text{main tasks}} + \underbrace{\mathcal{L}_{\text{aux}}}_{\text{auxiliary task}}
\end{equation}
FACL provides objective-oriented supervision for task tokens and helps prevent semantic drift.

\subsection{STARK: Semantic Tree Attention with Reorganized KV Cache}

Generative retrieval methods based on semantic IDs typically employ beam search for autoregressive decoding, expanding the top-$K$ candidate sequences along the batch dimension. However, this conventional implementation incurs severe efficiency bottlenecks. 

Specifically, the repetitive duplication and rearrangement of KV caches across the batch dimension incur significant memory bandwidth overhead, while independently processing these batched candidates leads to redundant attention computations over shared prefixes. These inefficiencies are further compounded by the extremely short length of semantic IDs, which causes mainstream attention kernels optimized for long sequences to suffer from severe under-utilization due to excessive padding.

To address these issues, we draw inspiration from tree attention in speculative decoding and propose STARK (Semantic Tree Attention with Reorganized KV Cache), an efficient inference scheme tailored to semantic ID decoding.
\begin{figure*}[!t]
    \centering
    \includegraphics[width=0.8\textwidth]{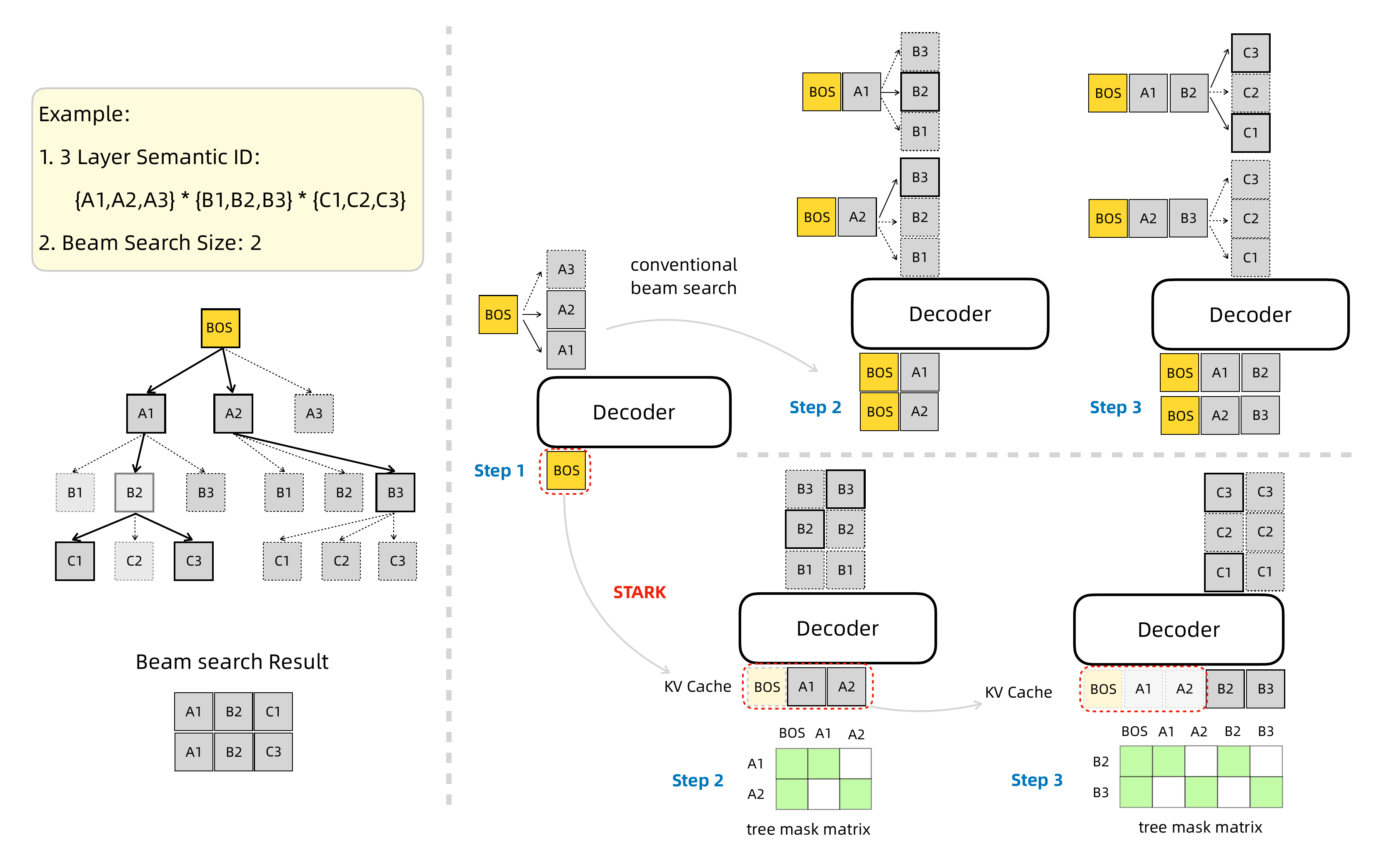}
    \caption{STARK inference mechanism. It uses a sequence-dimension tree attention mask to eliminate redundant prefix computations and data copying.}
    \label{fig:stark}
\end{figure*}

The core idea of STARK is to replace batch-dimension expansion with sequence-dimension expansion. At each decoding step, instead of copying the top-$K$ candidates into $K$ independent batch samples, we concatenate candidate tokens along the sequence dimension and use a predefined tree attention mask to control visibility. Each token can only attend to its ancestor nodes in the decoding tree and itself. This preserves the decoding results of standard beam search while improving efficiency.

Formally, let $P^{(l)}=\{p_1^{(l)},p_2^{(l)},\ldots,p_k^{(l)}\}$ denote the candidate paths at the $l$-th semantic level. STARK constructs a tree mask $\mathbf{M}^{(l)}$:
\begin{equation}
\mathbf{M}^{(l)}_{i,j} =
\begin{cases}
1, & \text{if } j \in \mathrm{Anc}(p_i^{(l)}) \text{ or } j=i,\\
0, & \text{otherwise},
\end{cases}
\end{equation}

\begin{table}[t]
    \centering
    \small
    \caption{End-to-end latency and throughput Comparision of STARK.}
    \label{tab:latency}
    \begin{tabular}{lcccc}
        \toprule
        \textbf{Method} & \textbf{Batchsize} & \textbf{QPS} & \textbf{AVG Lat(ms)} & \textbf{P99 Lat(ms)} \\
        \midrule
        w/o STARK & 1 & 119 & 30.9 & 33.5 \\
        STARK & 1 & 219 & 14.1 & 15.6 \\
        STARK & 8 & 596 & 26.1 & 26.8 \\
        \bottomrule
    \end{tabular}
\end{table}

where $\mathrm{Anc}(p_i^{(l)})$ denotes the ancestor positions of path $p_i^{(l)}$. This mask guarantees that candidate branches remain causally isolated while shared prefixes are computed exactly once.

STARK encodes the beam-search tree with a pre-computed attention mask. Each candidate token can only attend to its ancestor prefixes and itself, preserving causal isolation between different candidate paths. Since semantic ID depth and beam width are fixed at serving time, both the tree mask and KV cache remapping can be prepared as lightweight index operations rather than repeated data copies.

The KV cache organization follows the same tree topology. When two candidate paths share a prefix, the corresponding keys and values are stored only once and referenced by descendant nodes through the pre-computed index mapping. Therefore, candidate expansion primarily appends newly selected semantic tokens to the tree rather than duplicating the entire prefix cache for every beam. This property is especially beneficial for semantic ID generation, where the decoding length is short but beam width can be large.

Overall, STARK reduces data copying, avoids redundant prefix computation, and improves attention kernel utilization while preserving standard beam-search results. STARK achieves a 200\% throughput improvement in industrial-scale scenarios, supporting large-scale deployment of generative recommendation systems.

\section{Experiments}

\subsection{Setup}
\subsubsection{Dataset}
We evaluate \sysname on large-scale e-commerce recommendation logs from the Lazada homepage ``Guess You Like'' scenario. Multi-scenario pre-training uses mixed behavior data collected from multiple business scenarios, while scenario-specific alignment uses target-scenario logs with behavior labels. The evaluation assesses whether the ground-truth item, under a given behavior objective appears within the generated Top-$K$ candidate set, and whether the ranking module improves behavior prediction quality on the generated candidates.

Due to confidentiality reasons, we omit sensitive production statistics including user volume, item volume, and exact traffic scale. All compared methods are evaluated under the identical data split, semantic ID tokenizer, and serving constraints unless otherwise specified.

\subsubsection{Baseline Models}
We compare \sysname with several representative semantic ID-based generative recommendation baselines, including TIGER~\cite{rajput2023recommender}, OneRec~\cite{deng2025onerec}, and OneRec-V2~\cite{zhou2025onerecv2technicalreport}. These methods are closely related to our setting because they formulate recommendation as semantic ID generation.

\subsubsection{Metrics}
For generative retrieval, we adopt Hit Rate at top-$K$ (HR@$K$), where a hit is counted if the target item appears in the top-$K$ generated candidates.

In the multi-scenario pre-training stage, retrieval evaluation is conducted using click-based HR@$K$, since click behavior provides the main supervision signal at this stage. In scenario-specific alignment, we further report behavior-specific HR@$K$ under different target behaviors, including click (Clk-HR), add-to-cart (Atc-HR), and purchase (Pay-HR), to assess retrieval quality under multiple business objectives. For the ranking module, we additionally report AUC and GAUC for click, add-to-cart, and purchase prediction. We report HR@50, HR@100, HR@200, and HR@500 in retrieval experiments unless otherwise noted.

\subsubsection{Implementation details}
\sysname uses an embedding dimension of $d = 128$, adopts grouped-query attention (GQA) with every 4 query heads sharing one key-value head, and employs a sparse MoE architecture that activates only approximately 7\% of the total parameters per token. All baselines and variants use identical embeddings and hardware for fair comparison.

For generative retrieval inference, we decode the semantic ID sequence with beam search and set the beam width at each autoregressive step to $(512, 512, 1024)$ for the codebook layers. The wider beam on the last layer enlarges the hypothesis space when disambiguating among fine-grained candidates.

\subsection{Offline Performance}
We first evaluate the overall retrieval quality of \sysname against state-of-the-art baselines. Table~\ref{tab:offcompare} presents the offline comparison evaluated across multiple business scenarios. Under identical configurations, \sysname-M consistently surpasses all semantic ID-based baselines across all evaluated metrics. 
\begin{table}[!htbp]
    \centering
    \small
    \caption{Offline comparison of retrieval models.}
    \label{tab:offcompare}
    \begin{tabularx}{\columnwidth}{X cccc}
        \toprule
        \textbf{Model} & \textbf{HR@50} & \textbf{HR@100} & \textbf{HR@200} & \textbf{HR@500} \\
        \midrule
        TIGER & 0.1445 & 0.2026 & 0.2698 & 0.3675 \\
        OneRec-V2 & 0.1529 & 0.2126 & 0.2816 & 0.3812 \\
        OneRec & 0.1538 & 0.2151 & 0.2855 & 0.3866 \\
        \sysname-M (Ours) & \textbf{0.1579} & \textbf{0.2195} & \textbf{0.2896} & \textbf{0.3913} \\
        \bottomrule
    \end{tabularx}
\end{table}
\begin{table}[!htbp]
    \centering
    \scriptsize
    \caption{Behavior-specific hit rates of the UniSGR training paradigm evaluated on the Lazada homepage ``Guess You Like'' scenario.}
    \label{tab:main_results}
    \begin{tabular}{lcccccc}
        \toprule
        Model & \multicolumn{3}{c}{HR@100} & \multicolumn{3}{c}{HR@500} \\
        \cmidrule(lr){2-4} \cmidrule(lr){5-7}
        & Clk & Atc & Pay & Clk & Atc & Pay \\
        \midrule
        Pre-train Only & 0.1886 & 0.2276 & 0.2783 & 0.3893 & 0.4288 & 0.4768 \\
        Scenario Align Only & 0.2297 & 0.2397 & 0.2573 & 0.4334 & 0.4374 & 0.4386 \\
        Two-stage Training & \textbf{0.2842} & \textbf{0.3073} & \textbf{0.3408} & \textbf{0.5166} & \textbf{0.5369} & \textbf{0.5632} \\
        \bottomrule
    \end{tabular}
\end{table}

\subsection{Ablation Study}
We conduct comprehensive ablation studies on key components of \sysname. Specifically, we examine the impact of semantic ID design, model variants, scaling behavior across different model sizes, and the auxiliary task settings used in the ranking stage.

\subsubsection{Impact of Codebook Size}
To investigate the impact of codebook capacity, we evaluate symmetric 3-layer configurations ($L_1=L_2=L_3=K$) across varying sizes $K \in \{2048, 4096, 8192, 10240\}$, summarizing Clk-HR and collision rates in Table~\ref{tab:codebook_size}.

Increasing $K$ from 2048 to 8192 yields substantial gains, boosting Clk-HR@100 from 0.2917 to 0.3266. This improvement is primarily attributed by a sharp reduction in semantic collision rate within the codebook layers (from 45.00\% to 24.03\%), which enables better fine-grained item distinction during decoding.

However, further scaling $K$ to 10240 shows diminishing returns: Clk-HR remains essentially unchanged, and the collision rate drops only marginally to 23.58\%. This indicates that an excessively large codebook introduces parameter inefficiency and deep-layer under-utilization. Consequently, we adopt the $3L8192$ configuration as the optimal balance between representation capacity and retrieval accuracy.

\begin{table}[]
\centering
\small
\caption{Performance comparison of varying symmetric codebook sizes in the quantization module.}
\label{tab:codebook_size}
\begin{tabular}{lccc}
\toprule
\textbf{Model} & \textbf{HR@100} & \textbf{HR@500} & \textbf{Collision Rate} \\
\midrule
3L2048  & 0.2917 & 0.4051  & 45.00\% \\
3L4096  & 0.3126 & 0.4360  & 31.00\% \\
3L8192  & \textbf{0.3266} & 0.4597 & 24.03\% \\
3L10240 & 0.3264 & \textbf{0.4603} & \textbf{23.58\%} \\
\bottomrule
\end{tabular}
\end{table}

\subsubsection{Model Variants Ablation}
To understand the contribution of different architectural components in the \sysname decoder, we perform an ablation study by systematically removing key modules. The results are summarized in Table~\ref{tab:ablation}.

Removing the Mixture-of-Experts (MoE) routing mechanism (\textit{w/o MoE}) results in the pronounced performance degradation, with HR@100 dropping from 0.2195 to 0.1725. This underscores the critical role of sparse MoE in expanding model capacity and capturing diverse user interests without incurring prohibitive computational overhead. Replacing the SwiGLU activation with a standard feed-forward network (\textit{w/o SwiGLU}) leads to a noticeable decline across all metrics, confirming that SwiGLU affords superior non-linear expressiveness for semantic ID generation. 

\begin{table}[!htbp]
    \centering
    \small
    \caption{Ablation studies of the UniSGR retrieval module.}
    \label{tab:ablation}
    \begin{tabularx}{\columnwidth}{X cccc}
        \toprule
        \textbf{Model} & \textbf{HR@50} & \textbf{HR@100} & \textbf{HR@200} & \textbf{HR@500} \\
        \midrule
        \sysname-M & \textbf{0.1579} & \textbf{0.2195} & \textbf{0.2896} & \textbf{0.3913} \\
        w/o SwiGLU & 0.1523 & 0.2119 & 0.2813 & 0.3802 \\
        w/o MoE & 0.1229 & 0.1725 & 0.2313 & 0.3183 \\
        w/o Shared Expert & 0.1571 & 0.2180 & 0.2882 & 0.3883 \\
        \bottomrule
    \end{tabularx}
\end{table}

\begin{figure}[!htbp]
    \centering
    \begin{minipage}[t]{0.48\linewidth}
        \centering
        \includegraphics[width=\linewidth]{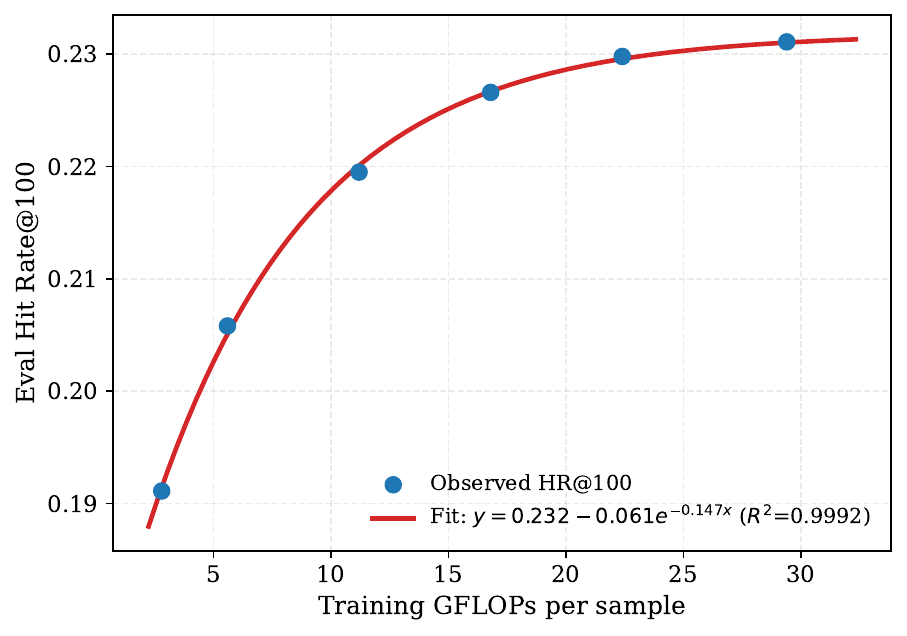}\\
        \small{(a) HR@100 scaling curve}
    \end{minipage}%
    \hfill
    \begin{minipage}[t]{0.48\linewidth}
        \centering
        \includegraphics[width=\linewidth]{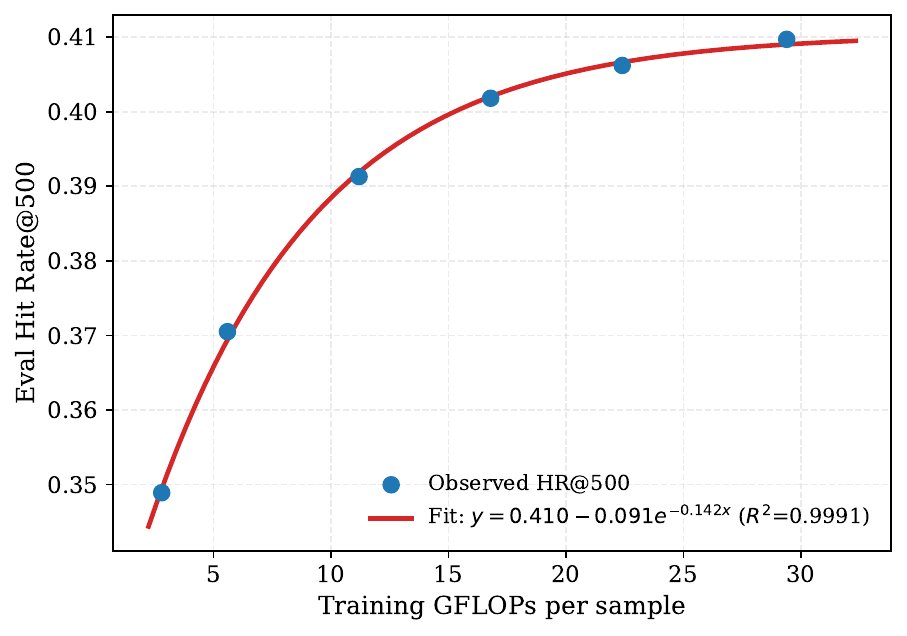}\\
        \small{(b) HR@500 scaling curve}
    \end{minipage}
    \caption{Scaling behavior of UniSGR across model sizes.}
    \label{fig:scaling}
\end{figure}
\subsubsection{Scaling}
We evaluate how UniSGR performs as the model size increases from 0.2B to 2.0B. As shown in Table~\ref{tab:scaling}, larger UniSGR models consistently achieve better results across all evaluation metrics, demonstrating clear and stable scaling behavior.

Specifically, compared to UniSGR-XS (0.2B), UniSGR-M (0.8B) improves HR@100 from 0.1911 to 0.2195 (+14.86\%) and HR@500 from 0.3489 to 0.3913 (+12.15\%). Further accuracy gains are observed when scaling to 1.2B, 1.6B, and 2.0B.

The marginal gains progressively diminish as the model size increases, suggesting that UniSGR benefits most from increased capacity in the small-to-medium scale regime, while exhibiting diminishing returns at larger scales, as also visualized in Figure~\ref{fig:scaling}.


\begin{table}[!htbp]
  \centering
  \small
  \caption{Performance scaling of UniSGR with varying model sizes.}
  \label{tab:scaling}
  \begin{tabularx}{\columnwidth}{X cccc}
    \toprule
    \textbf{Model} & \textbf{HR@50} & \textbf{HR@100} & \textbf{HR@200} & \textbf{HR@500} \\
    \midrule
    UniSGR-XS (0.2B) & 0.1364 & 0.1911 & 0.2549 & 0.3489 \\
    UniSGR-S (0.4B) & 0.1471 & 0.2058 & 0.2732 & 0.3705 \\
    UniSGR-M (0.8B) & 0.1579 & 0.2195 & 0.2896 & 0.3913 \\
    UniSGR-L (1.2B) & 0.1631 & 0.2266 & 0.2987 & 0.4018 \\
    UniSGR-XL (1.6B) & 0.1655 & 0.2298 & 0.3030 & 0.4062 \\
    UniSGR-XXL (2.0B) & 0.1664 & 0.2311 & 0.3048 & 0.4097 \\
    \bottomrule
  \end{tabularx}
\end{table}

\begin{table}[H]
    \centering
    \scriptsize
    \caption{Behavior-specific hit-rate performance on ranking module ablations.}
    \label{tab:rank_hitrate_ablation}
    \begin{tabular}{lcccccc}
        \toprule
        Model & \multicolumn{3}{c}{HR@100} & \multicolumn{3}{c}{HR@500} \\
        \cmidrule(lr){2-4} \cmidrule(lr){5-7}
        & Clk & Atc & Pay & Clk & Atc & Pay \\
        \midrule
        NTP & 0.2752 & 0.2841 & 0.3117 & 0.4962 & 0.5010 & 0.5117 \\
        VA-PMTP & 0.2777 & 0.3006 & 0.3382 & 0.5107 & 0.5287 & 0.5566 \\
        VA-PMTP + Ranking & 0.2903 & 0.3076 & 0.3621 & 0.5203 & 0.5380 & 0.5716 \\
        VA-PMTP + Ranking + TAT & 0.2924 & 0.3141 & 0.3614 & \textbf{0.5234} & \textbf{0.5389} & \textbf{0.5754} \\
        Full \sysname & \textbf{0.2932} & \textbf{0.3176} & \textbf{0.3636} & 0.5229 & 0.5379 & 0.5728 \\
        \bottomrule
    \end{tabular}
\end{table}

\begin{table}[H]
    \centering
    \scriptsize
    \caption{Ranking performance on scenario-specific alignment ablations.}
    \label{tab:rank_auc_ablation}
    \begin{tabularx}{\columnwidth}{Xcccccc}
        \toprule
        Model & \multicolumn{2}{c}{Click} & \multicolumn{2}{c}{ATC} & \multicolumn{2}{c}{Pay} \\
        \cmidrule(lr){2-3} \cmidrule(lr){4-5} \cmidrule(lr){6-7}
        & GAUC & AUC & GAUC & AUC & GAUC & AUC \\
        \midrule
        VA-PMTP & 0.5513 & 0.5684 & 0.5245 & 0.5360 & 0.5340 & 0.5603 \\
        VA-PMTP + Ranking & 0.5624 & 0.6218 & 0.5710 & 0.6780 & 0.5767 & 0.7662 \\
        VA-PMTP + Ranking + TAT & 0.5626 & 0.6219 & 0.5697 & 0.6855 & 0.5787 & 0.7656 \\
        Full \sysname & \textbf{0.5744} & \textbf{0.6334} & \textbf{0.5901} & \textbf{0.6978} & \textbf{0.6153} & \textbf{0.7870} \\
        \bottomrule
    \end{tabularx}
\end{table}

\subsubsection{Two-stage training}
To evaluate the two-stage training paradigm, we compare the full pipeline against both multi-scenario pre-training alone and scenario-specific alignment alone. As demonstrated in Table~\ref{tab:main_results}, the joint approach consistently outperforms these alternatives across Clk-HR, Atc-HR, and Pay-HR. This confirms the merit of performing scenario-specific alignment subsequent to a broad pre-training phase.

Furthermore, we use conventional NTP as the baseline and incrementally add the proposed alignment components. As shown in Table~\ref{tab:rank_hitrate_ablation}, VA-PMTP brings larger gains on add-to-cart and purchase than on click, indicating that value-aware generation is particularly helpful for higher-value behaviors.

On top of VA-PMTP, adding the ranking module improves both retrieval hit rate and ranking metrics. These results demonstrate the advantage of joint generation-ranking training: the ranking module provides direct business-objective supervision while sharing representations with the generator, as shown in Table~\ref{tab:rank_auc_ablation}.

TAT further improves consistency between generation and ranking by injecting objective-specific signals into the decoder. Reusing decoder states gives the ranking module richer interaction features, leading to the best GAUC and AUC results across click, add-to-cart, and purchase objectives.


These results also show that retrieval-oriented and ranking-oriented objectives are complementary rather than competing. VA-PMTP improves the generated candidate pool by making generation behavior-aware, while the ranking module improves the ordering and calibration of candidates under multiple business objectives. The consistent gains across both HR and AUC/GAUC suggest that sharing decoder representations does not merely add an auxiliary head; it changes the learned representation space in a way that benefits both candidate generation and final scoring.

\subsection{Online A/B Test}

To evaluate the online performance of \sysname, we conducted an online A/B test on the Lazada homepage ``Guess You Like'' recommendation scenario. The baseline is the production cascade recommendation system.

As shown in Table~\ref{tab:onlineperformance}, \sysname improves Item Page View (IPV) by 3.36\%, Transaction Count by 2.17\%, and Gross Merchandise Volume (GMV) by 5.68\% compared to the baseline. These results demonstrate that \sysname effectively enhances both user engagement and conversion metrics in a real-world industrial setting.

\begin{table}[H]
    \centering
    \small
    \caption{Absolute improvement of UniSGR compared to the baseline in the online A/B testing.}
    \label{tab:onlineperformance}
    \begin{tabular}{lccc}
         \toprule
         \textbf{Scenario} & \textbf{IPV} & \textbf{Transaction Count} & \textbf{GMV} \\
         \midrule
         Lazada Homepage & +3.36\% & +2.17\% & +5.68\% \\
         \bottomrule
    \end{tabular}
\end{table}

\section{Conclusion}
In this work, we propose \sysname, a unified framework integrating semantic ID generation and multi-objective ranking. To adress the inherent limitations of cascaded architectures, we introduce a two-stage training paradigm: multi-scenario pre-training followed by scenario-specific alignment. Within this paradigm, Value-Aware Parallel Multi-Token Prediction and Task-Aware Tokens guide candidate generation toward high-value business objectives. Additionally, we design STARK, a tree-attention inference strategy that accelerates decoding by eliminating redundant prefix computations. Extensive evaluations on a large-scale e-commerce platform validate the superiority and industrial scalability of \sysname.

\clearpage
\FloatBarrier
\printbibliography

\end{document}